\documentclass[journal]{IEEEtran}

\usepackage{cite}
\usepackage{bm}
\usepackage{amsmath}
\usepackage{amssymb}
\usepackage{color}
\usepackage[caption=false,font=footnotesize]{subfig}
\usepackage{graphicx}
\usepackage{float}
\usepackage{caption}
\usepackage{enumerate}
\usepackage{url}
\usepackage{soul}
\usepackage{soul}
\soulregister\cite7 
\soulregister\ref7 
\soulregister\pageref7

\ifCLASSINFOpdf
\else
\fi
\begin{document}
%
\title{An Unsupervised Deep Learning Method for Multi-coil Cine MRI}
%
%

\author{Ziwen~Ke, Jing~Cheng, ~\IEEEmembership{Student Member,~IEEE}, Leslie~Ying, ~\IEEEmembership{Senior Member,~IEEE}, Hairong~Zheng, ~\IEEEmembership{Senior Member,~IEEE}, Yanjie~Zhu*  and~Dong~Liang,~\IEEEmembership{Senior Member,~IEEE} 
\thanks{*Corresponding author: yj.zhu@siat.ac.cn}
\thanks{Yanjie Zhu, Jing Cheng, Hairong Zheng and Dong Liang are with Paul C. Lauterbur Research Center for Biomedical Imaging, Shenzhen Institutes of Advanced Technology, Chinese Academy of Sciences, Shenzhen, China}
\thanks{Ziwen Ke and Dong Liang are with Research Center for Medical AI, Shenzhen Institutes of Advanced Technology, Chinese Academy of Sciences, Shenzhen,
	China}
\thanks{Ziwen Ke and Jing Cheng are also with Shenzhen College of Advanced Technology, University of Chinese Academy
	of Sciences, Shenzhen,
	China}
\thanks{Leslie Ying is with Department of Biomedical Engineering and the Department of Electrical Engineering, The State University of New York, Buffalo, NY, USA}
\thanks{Code available at https://github.com/Keziwen/Unsupervised-via-TIS}
}


%
%

\markboth{Submitted to Physics in Medicine and Biology}
{Ziwen Ke \MakeLowercase{\textit{et al.}}: Unsupervised learning via time interleaved sampling}
%



\maketitle

\begin{abstract}
Deep learning has achieved good success in cardiac magnetic resonance imaging (MRI) reconstruction, in which convolutional neural networks (CNNs) learn a mapping from the undersampled k-space to the fully sampled images. Although these deep learning methods can improve the reconstruction quality compared with iterative methods without requiring complex parameter selection or lengthy reconstruction time, the following issues still need to be addressed: 1) all these methods are based on big data and require a large amount of fully sampled MRI data, which is always difficult to obtain for cardiac MRI; 2) the effect of coil correlation on reconstruction in deep learning methods for dynamic MR imaging has never been studied. In this paper, we propose an unsupervised deep learning method for multi-coil cine MRI via a time-interleaved sampling strategy. Specifically, a time-interleaved acquisition scheme is utilized to build a set of fully encoded reference data by directly merging the k-space data of adjacent time frames. Then these fully encoded data can be used to train a parallel network for reconstructing images of each coil separately. Finally, the images from each coil are combined via a CNN to implicitly explore the correlations between coils. The comparisons with classic k‐t FOCUSS, k‐t SLR, L+S and KLR methods on in vivo datasets show that our method can achieve improved reconstruction results in an extremely short amount of time.
\end{abstract}


%
\IEEEpeerreviewmaketitle

\section{Introduction}
%
%
%
%
\IEEEPARstart{C}{ardiac} 
 magnetic resonance (CMR) imaging is a noninvasive imaging technique that can be used to evaluate cardiac function and ventricular wall motion abnormalities. CMR imaging provides rich information for the clinical diagnosis of heart conditions \cite{finn2006cardiac}. However, cardiac motion adversely affects the quality of MR images and therefore limits the temporal and spatial resolution of cardiac MR imaging \cite{Liang2007Spatiotemporal}, especially in cardiac diseases such as tachycardia. Therefore, it is important to accelerate cardiac MR imaging without sacrificing image quality. 

Usually, the fast CMR approach requires a priori information to remove the aliasing artifacts caused by the violation of the Nyquist sampling theorem \cite{jerri1977shannon}. The advanced approach is compressed sensing (CS)/Low-rank(LR) \cite{baraniuk2007compressive, lustig2007sparse}. Numerous CS/LR-based methods have been proposed to accelerate cardiac MR imaging \cite{jung2007improved, tsao2003k, liang2012k, wang2013compressed, otazo2015low, lingala2011accelerated, Nakarmi2017A, ShettyBi2019bi}. For example, k-t FOCUSS \cite{jung2007improved} took advantage of the sparsity of x-f support to reconstruct x-f images from the undersampled k-t space. K-t ISD \cite{liang2012k} incorporated additional information on the support of the dynamic image in x-f space based on the theory of CS with partially known support. LSD \cite{wang2013compressed} employed a 3D patch-based spatiotemporal dictionary for sparse representations of dynamic image sequences. A typical low-rank example is L+S \cite{otazo2015low}, in which the nuclear norm was used to enforce low rank in L and the L1 norm was used to enforce sparsity in S. K-t SLR \cite{lingala2011accelerated} exploited the correlations in a dynamic imaging dataset by modeling the data to have a compact representation in the Karhunen-Louve transform (KLT) domain. KLR \cite{Nakarmi2017A} developed an algorithm with a kernel-based low-rank model that generalized the conventional low rank formulation. These methods greatly improved the spatiotemporal resolution of dynamic MR imaging; however, their iterative solution procedures require relatively long time to achieve high-quality reconstructions, and regularization parameter selection is empirical. Additionally, most of these CS-based approaches exploit a priori information only from the to-be-reconstructed images or from only a few reference images \cite{dong2019deep}: prior knowledge from big data is not utilized.

Recently, deep learning-based methods have been proposed and successfully applied to MR imaging \cite{wang2016accelerating, kwon2017parallel, han2018deep, zhu2018image, eo2018kiki, sun2018compressed, quan2018compressed, schlemper2018deep, qin2018convolutional, shan2019dimension, sun2016deep, hammernik2018learning, cheng2019modellearning, aggarwal2018modl, cheng2019model}. There are mainly two categories of deep learning-based fast MRI: (1) end-to-end learning methods \cite{wang2016accelerating, kwon2017parallel, han2018deep, zhu2018image, eo2018kiki, sun2018compressed, quan2018compressed} and (2) model-based unrolling methods \cite{sun2016deep, hammernik2018learning, cheng2019modellearning, aggarwal2018modl, schlemper2018deep, qin2018convolutional, shan2019dimension}. The end-to-end methods utilize bid data information to train a universal network for learning the mapping between the undersampled and fully sampled data pairs in an end-to-end manner. For example, in \cite{wang2016accelerating}, a plain convolutional neural network (CNN) was trained to learn the mapping relationship between undersampled brain MR images and fully sampled brain MR images. AUTOMAP \cite{zhu2018image} used a combination of fully connected networks and CNNs to learn the mapping from undersampled k-space to reconstructed image. The model-based methods unroll the optimization algorithm iterations to the neural network so that the network can automatically learn the hyperparameters or transformations in the optimization algorithm. Typical model-based networks include ADMM-Net \cite{sun2016deep}, VN-Net \cite{hammernik2018learning}, and learned PD \cite{cheng2019modellearning}. Model-based unrolling methods often achieve better reconstruction quality from less data than do end-to-end learning methods \cite{cheng2019model}. There are mainly three peer-reviewed deep learning works on dynamic MR imaging, namely, DC-CNN (A Deep Cascade of Convolutional Neural Networks) \cite{schlemper2018deep}, CRNN (Convolutional Recurrent Neural Networks) \cite{qin2018convolutional}, and DIMENSION (a DynamIc MR imaging method with both k-spacE aNd Spatial prior knowledge integrated via multI-supervised netwOrk traiNing) \cite{shan2019dimension}. DC-CNN proposed a deep cascade of convolutional neural networks to accelerate the data acquisition process by using data consistency layers. CRNN simultaneously learned the spatiotemporal dependencies of cardiac image series by exploiting bidirectional recurrent hidden connections across time sequences. DIMENSION developed a multi-supervised network training technique to simultaneously constrain both the frequency and the spatial domain information to improve the reconstruction accuracy. However, all three of these methods require a large amount of fully sampled cardiac MR images as the ground truth. Collections of these fully sampled images are always difficult, especially breath-holding and regular heart rhythms are required in the acquisition.  

In addition to a priori information, the spatial variance coil sensitivity provided by the phase array coil also plays an important role in fast MRI. Parallel MRI utilizes coil sensitivity to accelerate MR image acquisition and has been widely used in clinical scans. Deep learning methods based on parallel imaging have been studied \cite{8962951, hammernik2018learning, knoll2019assessment, aggarwal2018modl, kwon2017parallel, jun2019parallel, akccakaya2019scan
}. However, most CMR deep learning reconstruction studies utilize simulated single-channel k-space data to train the network, which leads to the underutilization of coil correlation. In other words, few of them have been applied to cardiac MR imaging. 

To solve the above issues, we propose an unsupervised deep learning framework for parallel cardiac MRI in this paper. A time-interleaved acquisition scheme is designed that can build a set of fully encoded reference data by merging all frames. These fully encoded data can be used to train a parallel network for reconstructing the image of each coil separately. A model-based reconstruction network (ADMM-Net-III \cite{cheng2019model}) developed by our group was employed in our study. Finally, the coil correlations are explored, and the coil images are combined via another CNN. Our contributions can be summarized as follows:
\begin{enumerate}
	\item We propose an unsupervised framework for dynamic MRI. In our framework, the acquisition of fully sampled data for network training is no longer needed, which is one of the greatest difficulties in deep learning-based cardiac imaging, especially breath-holding conditions and regular heart rhythms are required in the acquisition scheme. Specifically, a time-interleaved acquisition scheme was used, and the signals from all frames were merged to build a set of fully encoded reference data for network training. This is the first time that an unsupervised approach has been applied to dynamic MR imaging.
	\item Different from previous deep learning methods for dynamic MRI, which focus on single-channel MRI, we propose a deep learning-based strategy for multi-coil dynamic MRI. The proposed framework can explore coil correlations and decrease the complexity of network learning because coil images from multichannel data have simpler statistical distributions than do single-channel data. Although deep learning methods based on parallel imaging have been studied previously, few have focused on cardiac MR imaging. To the best of our knowledge, this is the first time that a deep learning network has been applied to multi-coil dynamic MRI.
	\item The experimental results show that the proposed method is superior to conventional CS-based methods such as k-t FOCUSS, k-t SLR, L+S and KLR, and it has a much shorter runtime. These findings demonstrate the effectiveness of unsupervised learning and parallel networks in cardiac MRI.
\end{enumerate}

The rest of this paper is organized as follows. Section II states the problem and the proposed methods. Section III summarizes the experimental details and the results to demonstrate the effectiveness of the proposed method. Finally, a discussion and conclusions are presented in Section IV and V, respectively.

\section{METHODOLOGY}
\subsection{Problem Formulation}
The goal of dynamic MR imaging is to estimate an unknown image from undersampled k-space data. Specifically, for 2D cardiac imaging, reconstruction is performed by solving the following optimization problem:
\begin{equation}
\label{eq_1}
\bm{d}^* = \arg\min_{\bm{d}} ||\bm{Ed}-\bm{m}||_2^2+\lambda\mathcal{R}(\bm{d})
\end{equation}
Here $\bm{E}=\bm{PFC}$ is the encoding operator, $\bm{C}$ denotes coil sensitivity maps, $\bm{F}$ is a Fourier transform
and $\bm{P}$ is a under-sampling matrix. For each frame, the image to be reconstructed is first multiplied by the coil sensitivity profiles and then Fourier transformed to k-space. Finally, the k-space data is undersampled by the undersampling mask. $\bm{d}\in \mathbb{C}^{N_xN_yN_t}$ is the 2D dynamic image series and $\bm{m}\in\mathbb{C}^{N_xN_yN_tN_c}$ is the undersampled multichannel k-space data. The first term is the data fidelity, which ensures that the k-space of the reconstructed data is consistent with the actual measurements. The second term is often referred to as a prior regularization, and $\mathcal{R}$ is a prior regularization of $\bm{d}$. $\lambda$ is a regularization parameter. In CS-based methods, $\mathcal{R}(\bm{d})$ is usually a sparse prior of $\bm{d}$ in the temporal dimension. 

In CNN-based methods, $\mathcal{R}(\bm{d})$ is a CNN prior, that forces $\bm{d}$ to match the network output:
\begin{equation}
\label{eq_2}
\bm{d}^* = \arg\min_{\bm{d}} ||\bm{Ed}-\bm{m}||_2^2+\lambda||\bm{d}-f_{CNN}(\bm{m}|\bm{\theta})||_2^2
\end{equation}
where $f_{CNN}(\bm{m}|\bm{\theta})$ is the network output under the parameters $\bm{\theta}$. The purpose of the network training process is to find the optimal parameters $\bm{\theta}^*$; after the network is trained, the network output $f_{CNN}(\bm{m}|\bm{\theta}^*)$ is the reconstruction we want. The data fidelity term is important to achieve high-quality reconstruction; therefore, data consistency (DC) layers are often introduced in CNN-based methods.

\subsection{The Proposed Unsupervised Framework}
The proposed unsupervised learning framework for cardiac MRI is shown in Fig.\ref{fig1}. To simplify the symbols, we omit the channel dimension in this section, and it should be noted that Fig.\ref{fig1} is specific to each coil. The entire unsupervised framework can be divided into three components:
\begin{figure}[htbp]
	\centerline{\includegraphics[width=1.0\linewidth]{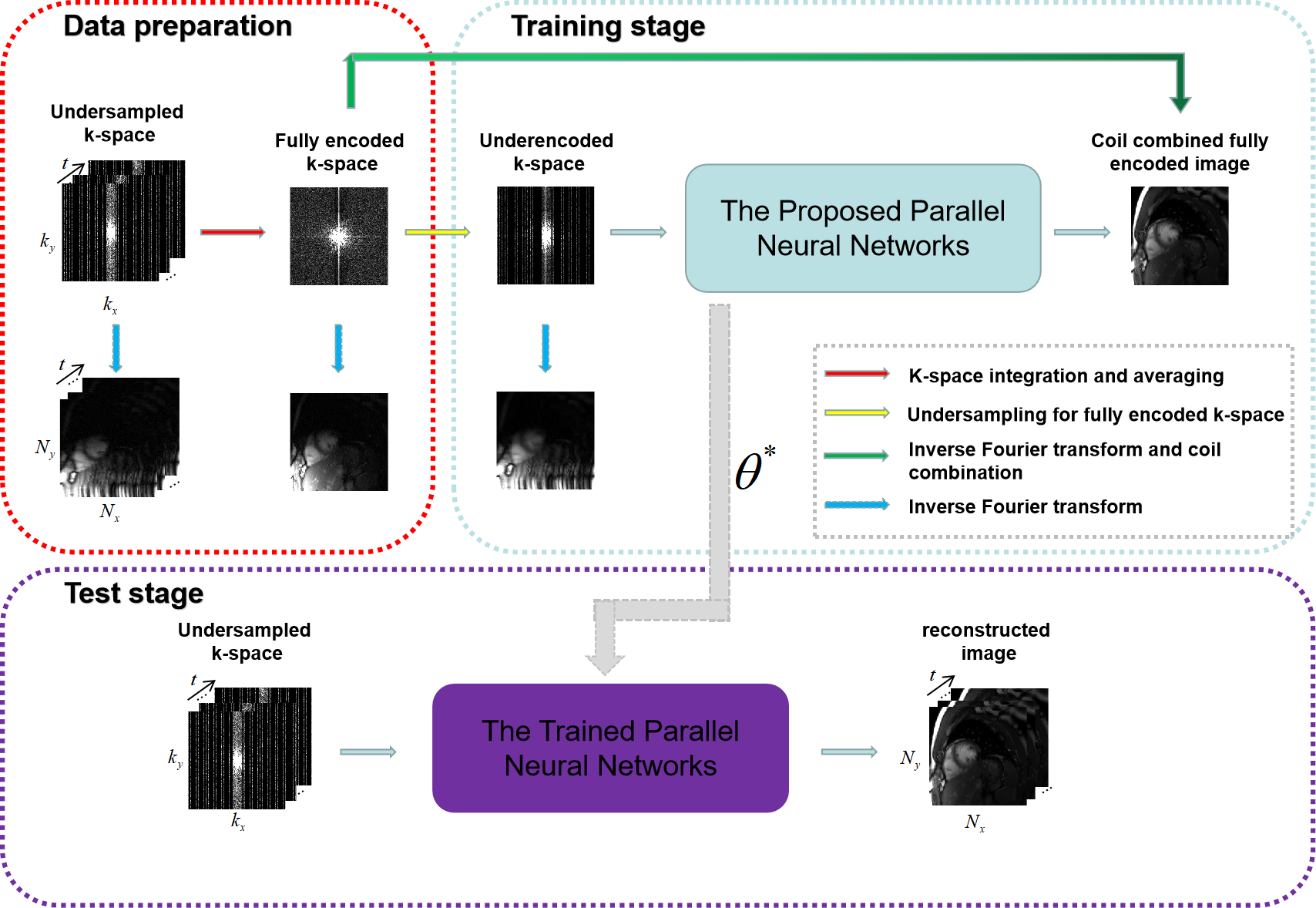}}
	\caption{The proposed unsupervised learning framework for dynamic MRI via time-interleaved sampling. Note that except for the network output in a coil-combined signal, all the signals are multichannel. During the data preparation stage, fully encoded data are built by directly merging adjacent time frames in a time-interleaved acquisition scheme. Then, the fully encoded data produce data pairs consisting of the network's input and its output for network training. In the test stage, in vivo undersampled k-space data are fed into the trained neural network to reconstruct the 2D dynamic MR images.\label{fig1}}
\end{figure}
\begin{enumerate}[-]
	\item \emph{Data preparation}: Undersampled k-space data are acquired according to a time-interleaved acquisition scheme. The time-interleaved acquisition scheme is shown in Fig.\ref{fig2}. All frames can be merged to build a complete set of k-space data by averaging, which is called the fully encoded k-space. In particular, we summed it in the temporal direction and divide by this count. After the fully encoded dataset is built, the network input and output data pairs can be obtained by retrospectively undersampling the fully encoded data with a designed sampling mask. Although in principle, the time-interleaved sampling scheme can be applied in uniform or random sampling patterns, for experimental convenience, we focus on uniform time-interleaved sampling. Although our framework is not limited to the number of merged frames, in the specific implementation, we merge all the frames of the undersampled k-space data to increase the SNR of the full-encoded data and to avoid GPU memory explosions. More implementation details can be found in Section III.A.
	\item \emph{Network training}: The proposed parallel neural network is described in the next section. The training datasets obtained above can be input into the network for network training. The network input consists of multichannel underencoded k-space data, and its output is the coil-combined fully encoded image. Although the training datasets are synthetic at this stage, they effectively represent real fully sampled data. More importantly, the temporal redundancies have been utilized through the construction of this dataset.
	\item \emph{Online test}: In the test stage, real in vivo undersampled k-space data are used as input to the trained network to reconstruct the 2D dynamic MR images. 
\end{enumerate}
\begin{figure}[htbp]
	\centerline{\includegraphics[width=1.0\linewidth]{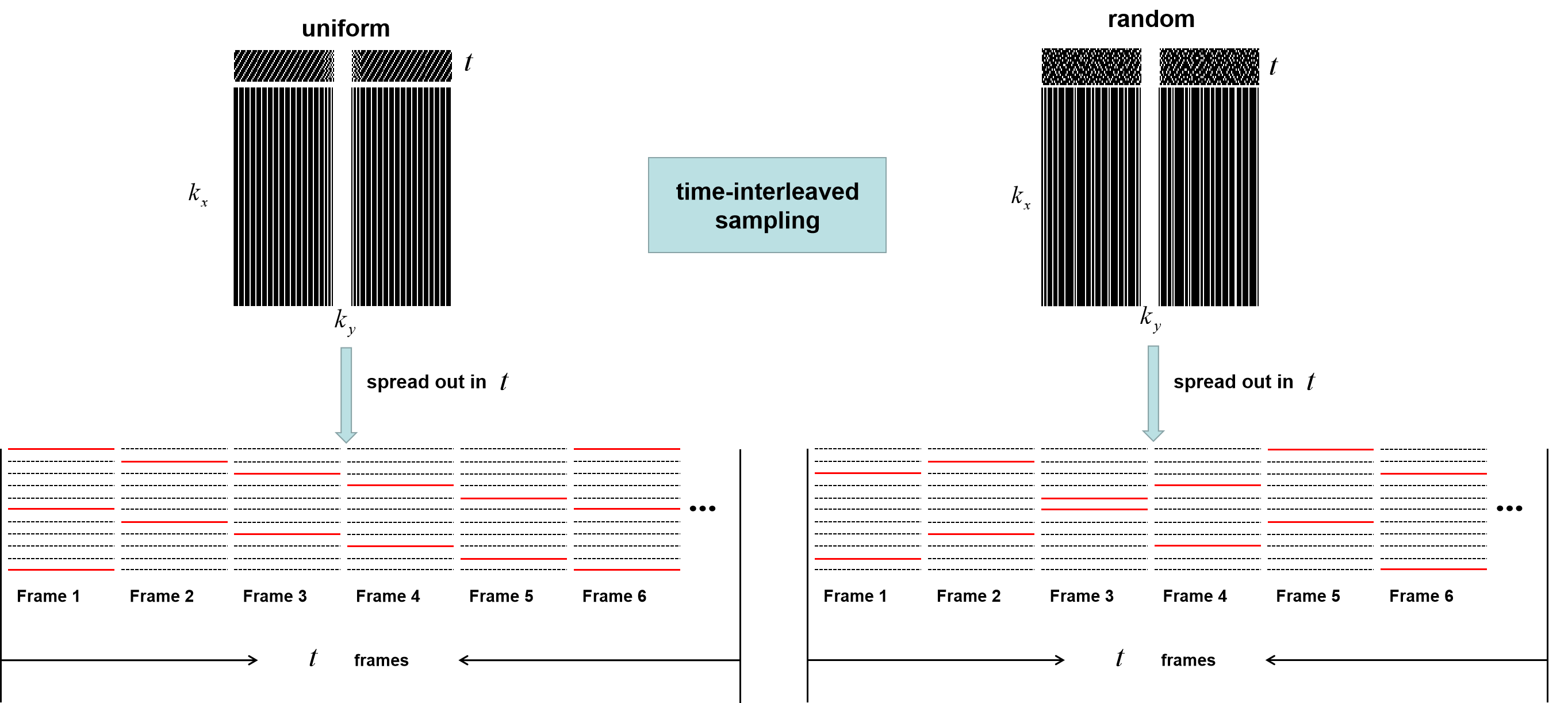}}
	\caption{The time-interleaved acquisition scheme. Two different undersampled patterns (uniform+random) at 5-fold acceleration acquired via a time-interleaved sampling scheme. In these two examples, at least five adjacent time frames need to be merged to build a complete set of k-space data. In general, more neighboring frames can be averaged to increase the SNR of the fully encoded data.\label{fig2}}
\end{figure}

The temporal resolution of the pseudo reference data is reduced. However, these reference data are used to obtain undersampled data for network training purposes. The merge operation is performed only during the training stage to establish the relationship from the undersampled data to the artifact-free image using the network. During the test stage, no test data merge operation is necessaty; therefore, the output image retains its original temporal resolution.

In summary, we use the time-interleaved sampling data to synthesize fully encoded data as references to achieve unsupervised learning. This framework has many advantages: (a) no fully sampled dynamic MR dataset is required; (b) coil correlations can be explored; (c) there is no need to explicitly calculate the coil sensitivity maps for coil combination; (d) the time redundancies are utilized. In Section III and IV, we demonstrate the effectiveness of this framework through abundant experiments.

\subsection{The Proposed Parallel Network}
\begin{figure*}[htbp]
	\centerline{\includegraphics[width=0.8\linewidth]{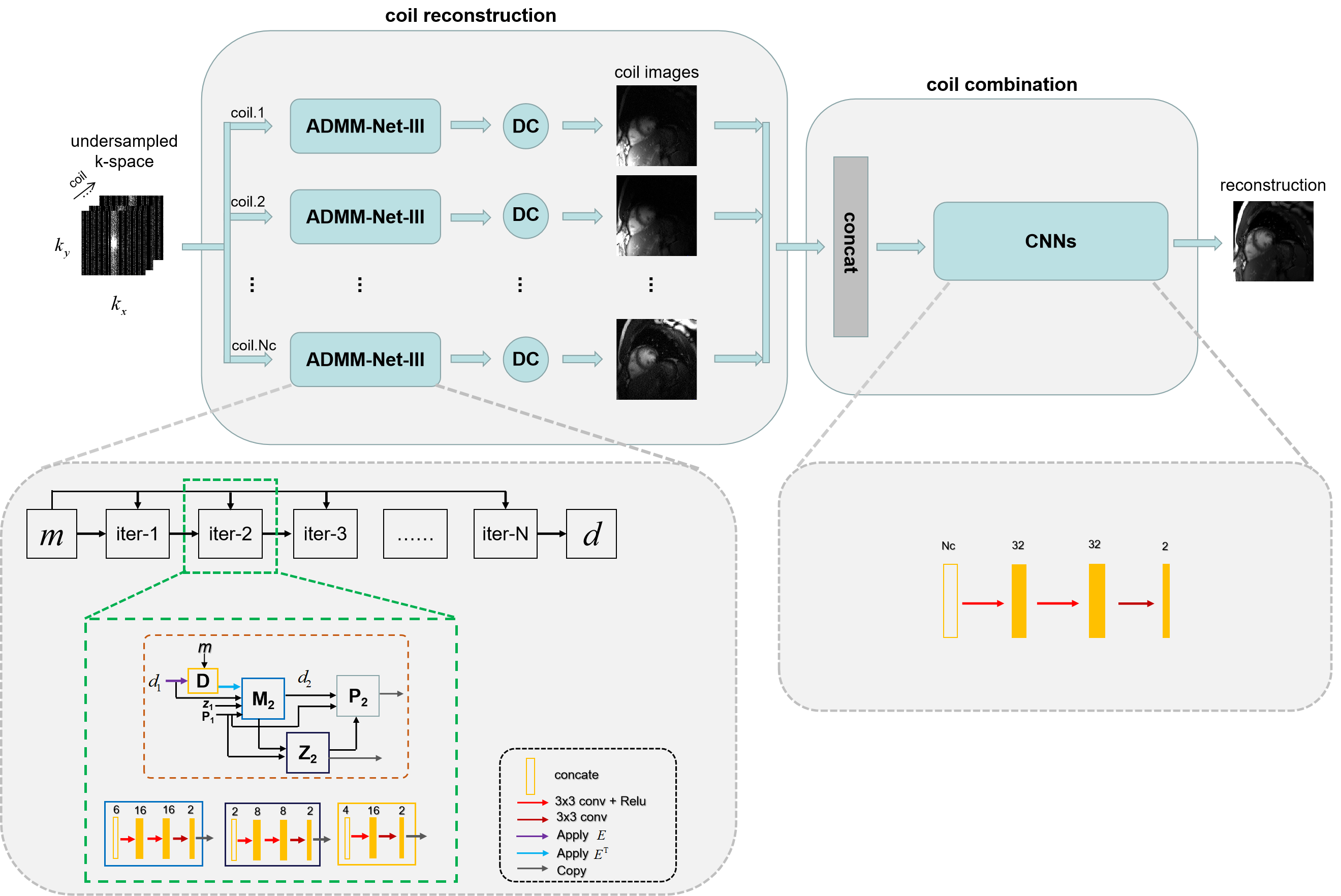}}
	\caption{The proposed parallel neural networks for MR reconstruction in a coil-by-coil manner.\label{fig3}}
\end{figure*}

 To complete the parallel reconstruction of underencoded multichannel k-space data, we propose a novel parallel neural network shown in Fig.\ref{fig3}. The proposed parallel network has two components: a reconstruction network to reconstruct each coil image, and a coil-combination network to explore coil correlations and combine all the coil images together. Specifically, the underencoded multichannel k-space data are fed into this parallel network. There is a separate network to reconstruct the coil data for each coil. We applied ADMM-Net-III \cite{cheng2019model} as the reconstruction network because it is a model-based unrolling method, which can obtain high-quality reconstruction results from less data. ADMM-Net-III is a generalized version of ADMM-Net \cite{sun2018compressed}. The iterations that ADMM-Net-III executes to reach a solution can be written as follows:
\begin{equation}
\label{eq_3}
\begin{cases}
D^{(n)}: \bm{\alpha}^{(n)}=\Gamma(\bm{Ed}^{(n-1)}, \bm{m}) \\
M^{(n)}: \bm{d}^{(n)}=\Pi(\bm{d}^{(n-1)}, \bm{z}^{(n-1)}-\bm{\beta}^{(n-1)}, \bm{E}^H\bm{\alpha}^{(n)}) \\
Z^{(n)}: \bm{z}^{(n)}=\Lambda(\bm{d}^{(n)}+\bm{\beta}^{(n-1)})\\
P^{(n)}: \bm{\beta}^{(n)}=\bm{\beta}^{(n-1)}+\tilde{\eta}(\bm{d}^{(n)}-\bm{z}^{(n)})
\end{cases}
\end{equation}
In ADMM-Net-III, the operators $\Gamma$, $\Pi$, $\Lambda$ and the parameter $\tilde{\eta}$ are all learned by the network. In contrast, only the priori regularization and the parameter are learned in the original ADMM-Net \cite{sun2016deep}. The operator $\Gamma$ refers to the function that corresponds to the deviation of data consistency, which is accomplished by the neural network. The operator $\Pi$ refers to the regularization term learned by the neural network. The operator $\Lambda$ refers to the auxiliary variables learned by the neural network. An in-depth explanation of ADMM-Net-III is beyond the scope of this article; however, the generalization process for ADMM-Net-III and its implementation results can be found in \cite{cheng2019model}. Unlike the original ADMM-Net-III model, each ADMM-Net-III model implemented for this study is followed by a data consistency (DC) layer because our previous exploration \cite{shanshan2019investigation} showed that the data consistency layer effectively improves the reconstruction quality. The DC is a backfill operation on the k-space: for the k-space coefficients that are initially unknown, we use the reconstructed values from the CNN. For the coefficients that have already been sampled, we correct the network predicted k-space with a combination of the actual sampled k-space and the predicted k-space \cite{shan2019dimension}. Then, all of the network reconstructed coil images are concatenated along the coil dimension and fed into another CNN that explores the coil correlations and implements a coil combination.

\begin{figure}[htbp]
	\centerline{\includegraphics[width=1.0\linewidth]{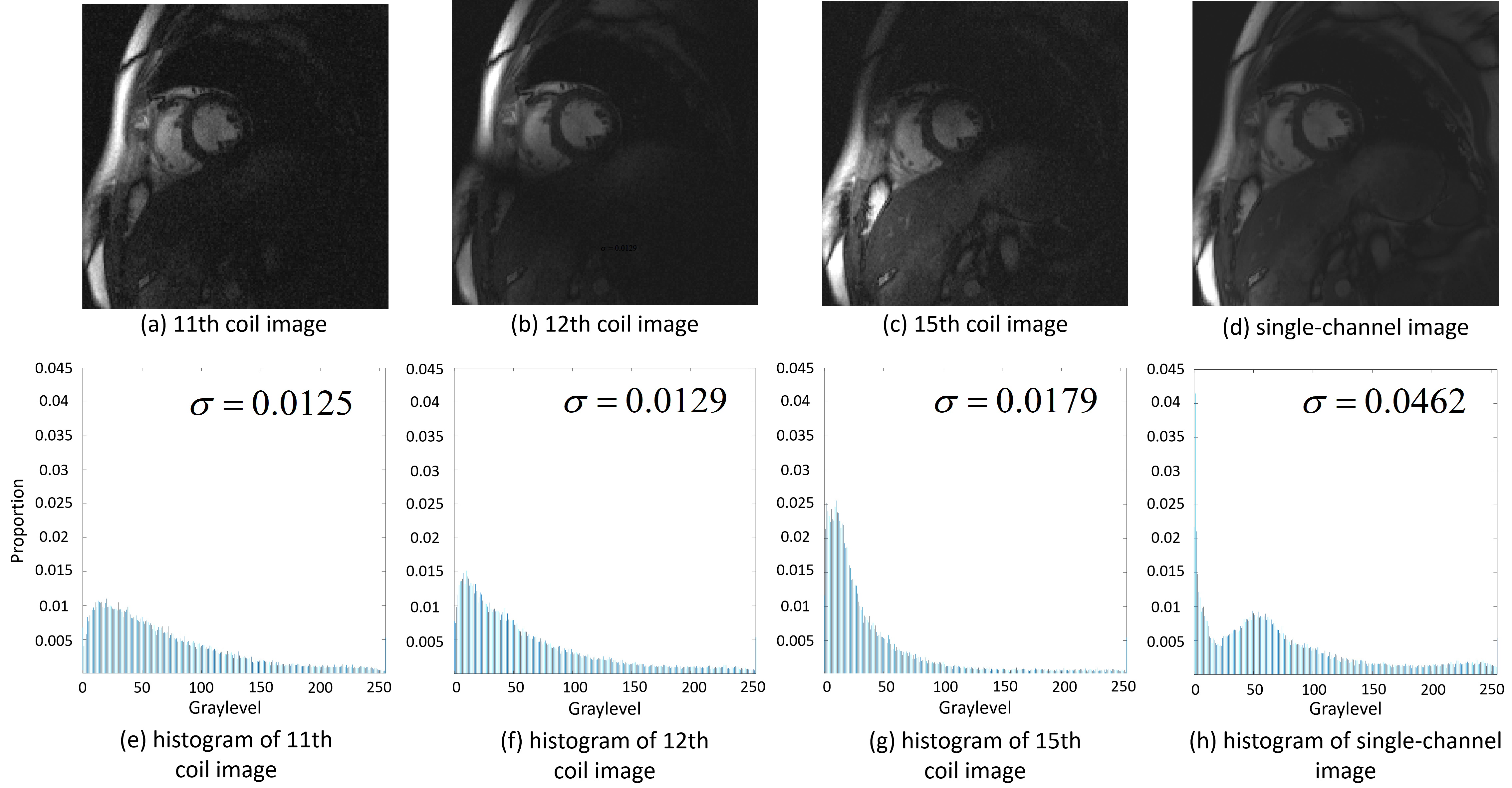}}
	\caption{The histograms of three coil images from multi-channel data and a single-channel image. From (a) to (d) are the 11th coil image, the 12th coil image, the 15th coil image and the single-channel image obtained by directly calculating the sum of squares (sos), while (e) to (h) \\hl{show} histograms of these images. The data are normalized to [0, 255] for convenient display. The second moment, $\sigma$, of each image is also given.\label{fig4}}
\end{figure}

Unlike other methods \cite{schlemper2018deep, qin2018convolutional, shan2019dimension} that use single-channel signals as input and output, our parallel network focuses on a multichannel scenario and could explore coil correlations. Another advantage of a multichannel approach to coil images is that single-channel data have a complicated distribution \cite{Fan_2018_ECCV}, which undoubtedly increases the difficulty of network learning. We visualized the statistical distributions of the single-channel and multichannel coil images; statistical histograms of both are provided in Fig.\ref{fig4}. The second moment, $\sigma$, which can measure the complexity of pixel histograms \cite{el2014content}, is also given. Its calculation formula is as follows:
\begin{equation}
\label{eq_second_moment}
\begin{cases}
\bm{\sigma}=\sqrt{\frac{1}{N}\sum_{i=1}^{N}(\bm{p_i}-\bm{\mu)}} \\
 \bm{\mu}=\frac{1}{N}\sum_{i=1}^{N}\bm{p_i}
\end{cases}
\end{equation}
where $\bm{p_i}$ is the value of the pixel $i$, and $N$ is the number of pixels in the image. A larger $\bm{\sigma}$ indicates a more complex image. To fairly compare the second moments of these images, all images were normalized in our experiments. The statistical histogram and quantitative indicators show that each coil image from the multi-channel data has a simpler statistical distribution than does the combined single-channel image. In Section IV.A, we compare a single-channel model with a multichannel model under the proposed unsupervised framework and find that the multichannel model achieves better reconstruction results. 

\section{EXPERIMENTAL RESULTS}

\subsection{Setup}
\subsubsection{Data acquisition}
We collected 386 2D dynamic (2Dt) fully sampled cardiac MR data from 30 healthy volunteers using a 3T scanner (SIEMENS MAGNETOM Trio) with a balanced steady-state free precession (bSSFP) sequence. All the in vivo experiments were approved by the Institutional Review Board (IRB) of the Shenzhen Institutes of Advanced Technology under the accepted ID SIAT-IRB-190315-H0323, and informed consent
was obtained from each volunteer prior to beginning the experiments. Each scan contains a single-slice bSSFP acquisition with 25 temporal frames. Retrospective electrocardiogram ECG-gated segmented imaging was conducted, and each slice was acquired during one breath-hold of 15-20 sec. The following parameters were used for the bSSFP scans: FOV $330\times330$ mm, acquisition matrix $256\times256$, slice thickness = 6 mm, TR/TE = 3.0 ms/1.5 ms and 20 receiving coils. The temporal resolution was 40.0 ms. We randomly selected 25 volunteers for training and used the rest for testing. Deep learning typically requires large amounts of data for training \cite{lecun2015deep}; therefore, some data augmentation strategies were applied. The data augmentation pattern that we chose was rigid transformation-shearing. We sheared the original multichannel images along the x, y and t directions. The sheared size was $192\times192\times16$ ($x\times y\times t$), and the stride in the three directions was 12, 12 and 5. Finally, we obtained 2149 2Dt multichannel cardiac MR data of size $192\times192\times16\times20$ ($x\times y\times t\times coil$) for data preparation and training and 603 data items for testing. Although fully sampled k-space data are available in our acquisition, they are not seen by the proposed unsupervised learning framework; they are used only to obtain the undersampled data retrospectively.

In this work, we focus on the uniform time-interleaved sampling pattern shown in Fig.\ref{fig2} (left). We retrospectively undersampled each original fully sampled k-space data accroding to the uniform time-interleaved sampling pattern. Specifically, we fully sampled the frequency-encodes (along $k_x$) and uniformly undersampled the phase encodes (along $k_y$). Wherein, 16 central phase-encodes were ensured to be sampled. Finally, we obtained the undersampled k-space data for data preparation. In the data preparation stage, we merged all the frames of the undersampled k-space data rather than the adjacent frames to obtain high-SNR fully encoded training data. This approach brought additional benefits, such as the elimination of temporal redundancy and reduced the requirement of GPU memory. The merged reference data have a low temporal resolution. However, this merge operation is conducted only during the training stage to establish the relationship from the undersampled data to the artifact-free image using the network. In the testing stage, no merge operation occurs; therefore, the output image remains at its original temporal resolution.

For quantitative evaluation, we adopted mean square error (MSE), peak signal-to-noise ratio (PSNR) and the structural similarity index (SSIM).

\subsubsection{Network training and testing}
For network training, we divided each data into two channels that store the real and imaginary parts of the data, respectively. Therefore, the network input consists of underencoded multichannel k-space data $\mathbb{R}^{2N_xN_yN_c}$, and its outputs the coil-combined reconstructed images $\mathbb{R}^{2N_xN_y}$. The hyperparameters in the network were set as follows: for each ADMM-Net-III, the number of iterations was set to $N = 8$, and the number of convolution kernels was set as shown in Fig.\ref{fig3}; the size of each convolution kernel is $3\times3$. Xavier initialization \cite{glorot2010understanding} was used to initialize the network weights. The rectifier linear units (ReLU) \cite{glorot2011deep} function was selected as the nonlinear activation function. The minibatch size was 4. The exponential decay learning rate \cite{zeiler2012adadelta} was used in all the CNN-based experiments: the initial learning rate was set to 0.001 with a decay of 0.98. The loss function used in this work was mean squared error (MSE). All the models were trained using the Adam optimizer \cite{kingma2014adam} with parameters $\beta^1=0.9$, $\beta^2=0.999$ and $\epsilon=10^{-8}$. The merged k-space data were used only for training. During testing, the inputs were the undersampled k-space data from all frames, and the outputs were the reconstructed dynamic images.

The models were implemented on an Ubuntu 16.04 LTS (64-bit) operating system equipped with an Intel Xeon E5-2640 central processing unit (CPU) and Tesla TITAN Xp graphics processing unit (GPU, 12 GB memory) in the open framework TensorFlow \cite{abadi2016tensorflow} with CUDA and CUDNN support. Network training required approximately 56 hours and 100 epochs.

\subsubsection{Model configuration}
 Our proposed unsupervised framework supports many alternative options, for example: (1) in the data preparation stage, the time-interleaved sampling scheme is not limited to uniform or random sampling patterns; (2) in the training stage, the retrospectively undersampling mask is not limited to uniform or random sampling patterns;  and (3) in the test stage, the trained model has good generalizability to other sampling patterns. Discussing all the cases regarding which sampling patterns can be used in the three stages is beyond the scope of this article. Therefore, without loss of generality, for experimental convenience, we experimented only on typical cases. The model configurations (sampling patterns and acceleration) are arranged in TABLE \ref{models_configuration}. The experiments from Section III.B to Section IV.D are designed according to the Table 1. A 1D Gaussian random undersampling pattern \cite{jung2007improved}, which is one of the most common protocols in CS/LR-based methods, was applied in this paper. Specifically, we fully sampled the frequency encodes (along $k_x$) and randomly undersampled the phase encodes (along $k_y$) according to a zero-mean Gaussian variable density function. 
\begin{table}[!t]
	\renewcommand{\arraystretch}{1.1}
	\renewcommand\tabcolsep{3pt}
	\caption{\textcolor{black}{The model configuration of each section.}}
	\label{models_configuration}
	\centering
	\textcolor{black}{\begin{tabular}{|c|c|c|c|c|} \hline
			& Time-interleaved & Training & Testing & Acceleration \\ \hline
			Section III.B & Uniform &  Random & Random & 4/8 \\
			\hline
			Section IV.A & Uniform & Random & Random & 4\\
			\hline
			Section IV.B & Uniform & Random & Random & 4\\
			\hline
			Section IV.C & Uniform & Random & Random & 4\\
			\hline
			Section IV.D & Uniform & Uniform & Random/Uniform & 4\\
			\hline
	\end{tabular}}
\end{table}

\subsection{Comparisons to the State-of-the-art Methods}
\begin{figure*}[htbp]
	\centerline{\includegraphics[width=1.0\linewidth]{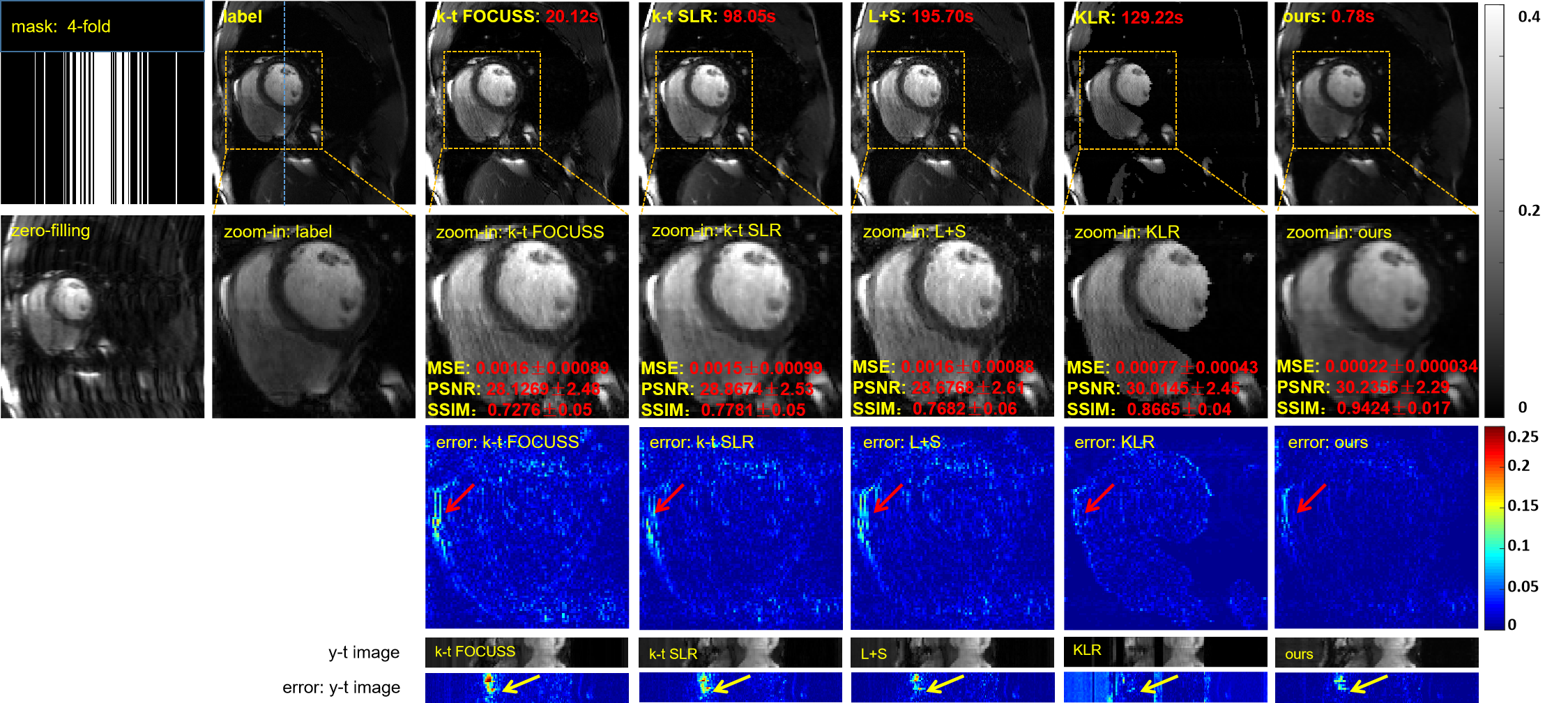}}
	\caption{The reconstruction results of the different methods (k-t FOCUSS, k-t SLR, L+S, KLR and the proposed method) at 4-fold acceleration. The proposed model is trained under a 1D random mask. The first row, from left to right, are the sampling mask, ground truth, and the reconstruction result of these methods. The second row, from left to right, shows the zero-filling and the enlarged view of their respective heart regions framed by a yellow box. The third row shows the error map (display ranges [0, 0.25]). The y-t image (extraction of the 124th slice along the y and temporal dimensions) and the error of the y-t image are also provided for each signal to show the reconstruction performance in the temporal dimension.\label{fig6}}
\end{figure*}

To demonstrate the efficacy of the proposed unsupervised learning method, we compared it with several state-of-the-art CS/LR methods including k-t FOCUSS \cite{jung2007improved}, k-t SLR \cite{lingala2011accelerated}, L+S \cite{otazo2015low} and KLR \cite{Nakarmi2017A}. We adjusted the parameters of the competing methods to elicit their best performance. A 1D random Gaussian mask was used for training and testing. 
The reconstruction results of these methods at 4-fold acceleration are shown in Fig.\ref{fig6}. The reconstruction results of the four CS-based methods contain fewer structural details and more artifacts than do the reconstruction results of the proposed method. We also enlarged the cardiac region and its error map for demonstration to show that our method achieves the best reconstruction performance in the cardiac region, especially the details marked by the red arrow. The y-t image and its error map, which were extracted from the 124th slice along the y and temporal dimensions, also clearly illustrate the superior performance of the proposed method. The evaluation indexes, MSE, PSNR, and SSIM, can be found in the enlarged view of the heart regions. All the quantitative results shown are the averaged results on the test data set, and the standard deviations are also given. We observe that the MSE, PSNR and SSIM indexes of the proposed method are the best among all the methods. The red numbers represent the reconstruction time of these methods for the entire volume. Our method also has the shortest reconstruction time--hundreds of times shorter than those of the other methods. 

\begin{figure*}[htbp]
	\centerline{\includegraphics[width=1.0\linewidth]{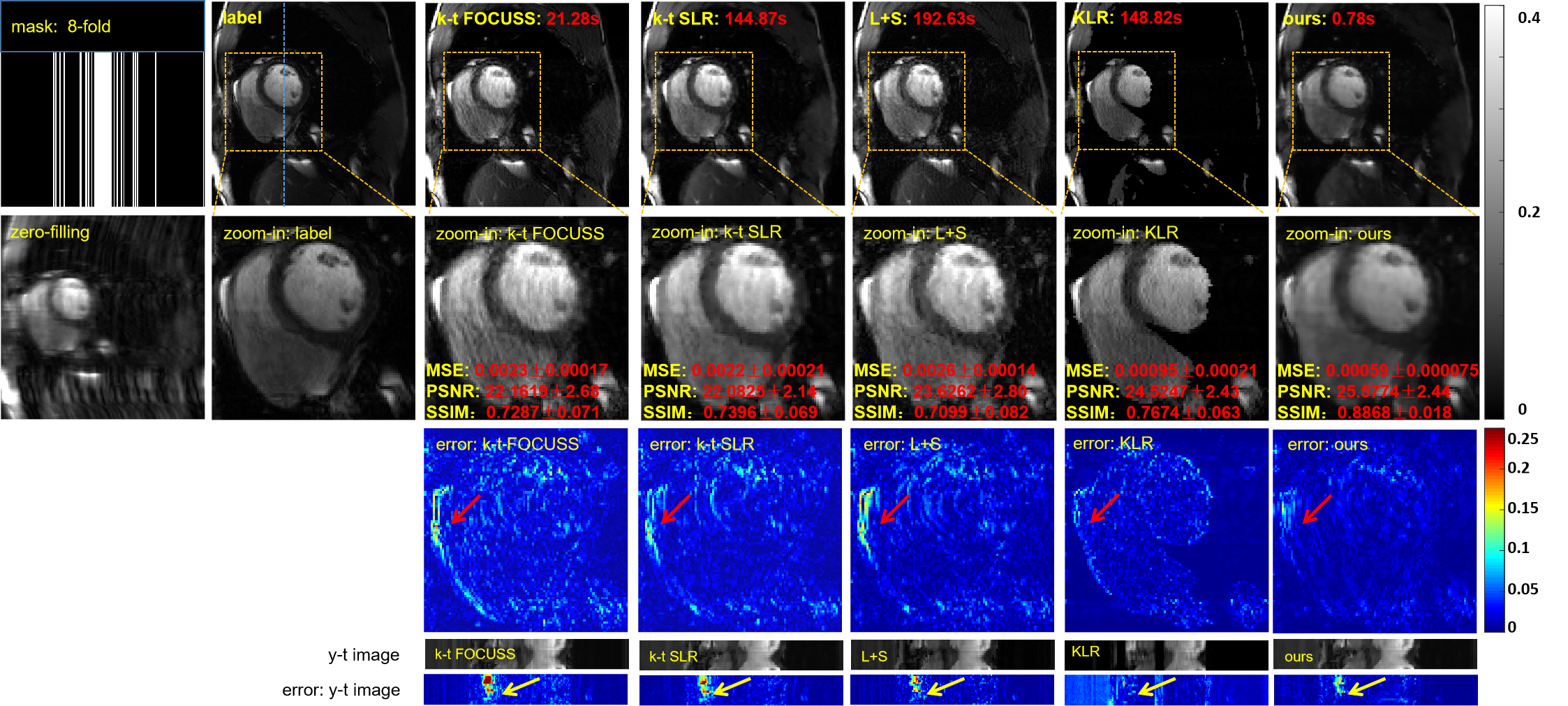}}
	\caption{The reconstruction results of the different methods (k-t FOCUSS, k-t SLR, L+S, KLR and the proposed method) at 8-fold acceleration. The proposed model is trained under a 1D random mask. The first row, from left to right, are the sampling mask, ground truth, and the reconstruction result of these methods. The second row shows, from left to right, the zero-filling and the enlarged view of the respective heart regions framed by a yellow box. The third row shows the error map (display ranges [0, 0.25]). The y-t image (extraction of the 124th slice along the y and temporal dimensions) and the error of the y-t image are also provided for each signal to show the reconstruction performance in the temporal dimension.\label{fig7}}
\end{figure*}

The reconstruction results of the different methods at 8-fold acceleration are shown in Fig.\ref{fig7}. At 8-fold acceleration, we can reach the same conclusion as with 4-fold acceleration.

\section{DISCUSSION}
\subsection{Single-channel Model versus Multichannel Model}
Currently, all three deep learning methods \cite{schlemper2018deep, qin2018convolutional,shan2019dimension} for cardiac MRI use single-channel data for network training and testing. In this section, we refer to these methods as single-channel methods and correspondingly refer to the proposed parallel imaging method as the multichannel method. We will explore whether the multichannel method exhibits superior reconstruction performance. To ensure a fair comparison, we explored the effectiveness of the single-channel methods and the multichannel method based on the same network structure (ADMM-Net-III) under the unsupervised framework proposed in this paper. Although we show only the unsupervised scheme in the multichannel case in Fig.\ref{fig1}, this unsupervised scheme can be conveniently changed to a single-channel case because the operations in Fig.\ref{fig1} focus on the temporal dimension and have nothing to do with the coil dimension. The two models differ in only two respects. First, the raw materials for data preparation are different: one consists of multichannel fully sampled k-space data, while the other consists of single-channel fully sampled k-space data by adaptively combining the above multichannel k-space data \cite{walsh2000adaptive}. Second, the single-channel model does not require a network to combine the coils; therefore it includes only the reconstruction part. In addition, to prove that the superiority of the multichannel model over the single-channel model is due to the ability of the multichannel model to learn the correlations between the coils rather than the model's capacity, we built a new single-channel model with the same network structure as the multichannel model we proposed, but for which the input data are multiple copies of the single-channel data. We call this model the single-channel-copied model.

\begin{figure*}[htbp]
	\centerline{\includegraphics[width=0.8\linewidth]{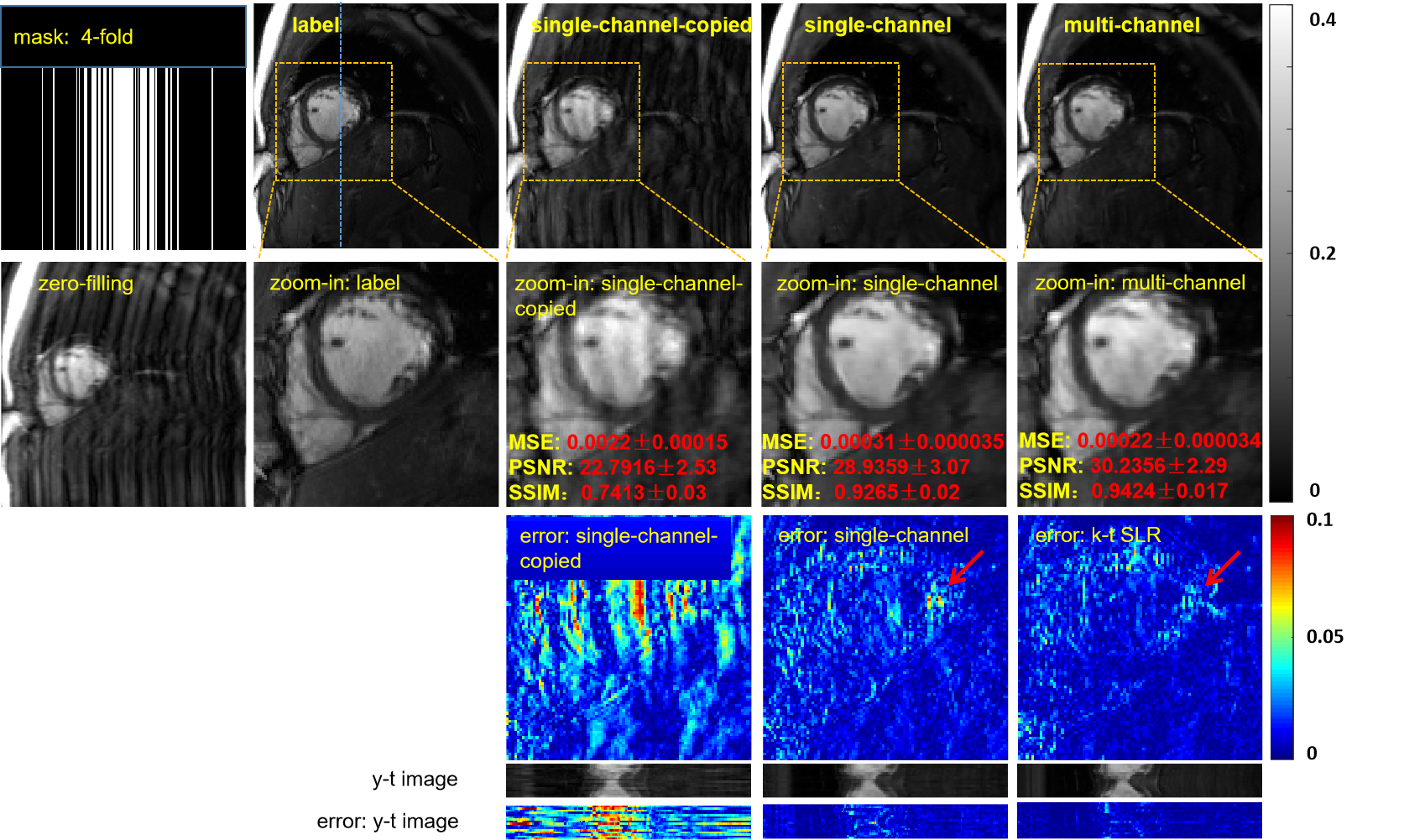}}
	\caption{The reconstruction results of the single-channel-copied, single-channel and multichannel models at 4-fold acceleration.  From left to right, the first row shows the sampling mask, the ground truth, and the reconstruction result of these three models. The second row shows, from left to right, the zero-filling and the enlarged view of their respective heart regions framed by a yellow box. The third row shows the error map (display ranges [0, 0.10]). The y-t image (extraction of the 124th slice along the y and temporal dimensions) and the error of the y-t image are also provided for each signal to show the reconstruction performance in the temporal dimension.\label{fig5}}
\end{figure*}

We trained these three models with a 1D random Gaussian sampling mask at 4-fold acceleration. The reconstruction results of the three models in Fig.\ref{fig5} clearly shows that the multichannel model restores more details (as indicated by red arrows) than does our single-channel model. The single-channel model not only loses more detail than the multichannel model, but also the reconstruction results are blurrier. The single-channel-copied model failed to reconstruct the dynamic MR image, despite having the same network structure as the multichannel model. Therefore, we can conclude that the proposed multichannel model achieves good reconstruction results because it explores the coil correlation well. The reasons why the dynamic MR image cannot be reconstructed as well in the single-channel-copied model may be as follows: (1) Each coil network learns the same features, resulting in an overspecificity of features, which reduces model generalizability; and (2) the errors or noise in each feature from each coil network have a superposition effect, resulting in large reconstruction errors. 

\subsection{Coil Reconstruction Network Options}
\begin{figure}[htbp]
	\centerline{\includegraphics[width=1.0\linewidth]{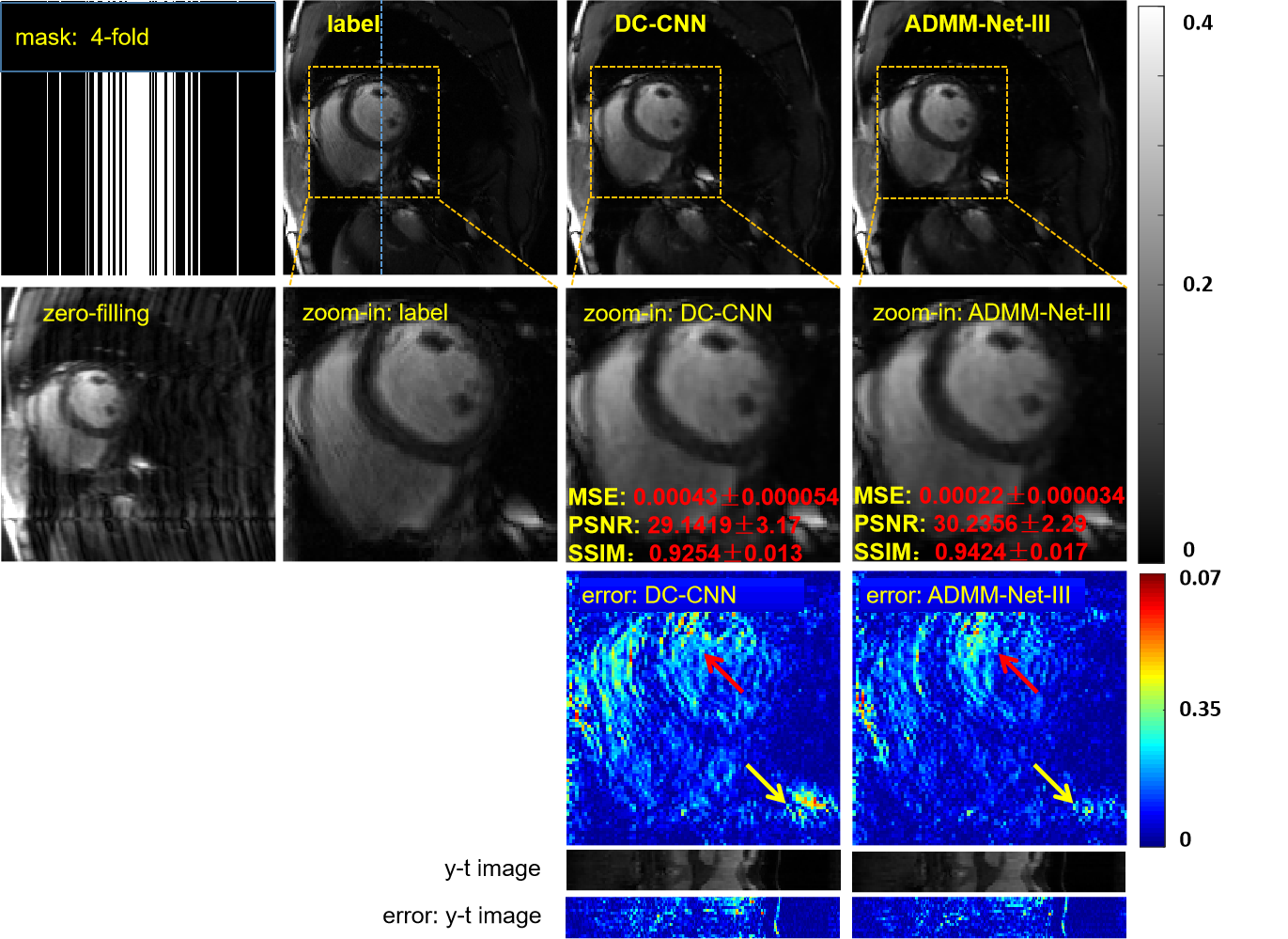}}
	\caption{Cardiac MR reconstruction results under the proposed framework with DC-CNN/ADMM-Net-III at 4-fold acceleration. The first row, from left to right, are the sampling mask, the ground truth, and the reconstruction results of DC-CNN and ADMM-Net-III. The second row shows, from left to right, the zero-filling and the enlarged view of the respective heart regions framed by a yellow box. The third row shows the error map (display range [0, 0.07]). The y-t image (extraction of the 124th slice along the y and temporal dimensions) and the error of the y-t image are also provided for each signal to show the reconstruction performance in the temporal dimension.\label{fig12}}
\end{figure}

Although we introduced the proposed parallel network in Section II.C, where ADMM-Net-III was selected as the reconstruction network, the full unsupervised framework is not limited to this specific reconstruction network. The reason we chose ADMM-Net-III is that it is a deep learning model-based unrolling method, which requires less data than do vanilla end-to-end methods and usually exhibits superior reconstruction performance compared to other methods. In this section, we compared the reconstruction results of ADMM-Net-III with those of DC-CNN \cite{schlemper2018deep}, which is a state-of-the-art deep learning method for dynamic MRI. We focused on a D5C5 model, which works well for the DC-CNN model and consists of five blocks (C5), each of which contains five convolutional layers (D5). We trained two models under the proposed unsupervised framework. The configuration parameters of the network remain unchanged. The only difference was that one reconstruction network used ADMM-Net-III and the other used D5C5. The two models had similar numbers of learned parameters: ADMM-Net-III had 762k parameters and D5C5 had 776k parameters. The reconstruction results are shown in Fig.\ref{fig12}. Both the qualitative and quantitative results have consistent observations: the coil reconstruction network using the ADMM-Net-III model has somewhat smaller artifacts and more details, especially in the heart region (as indicatd by the red and yellow arrows in the error maps). To create clearer comparisons, the display range was narrowed to [0, 0.07].

\subsection{The Necessity of the Coil Combination Network}
\begin{figure}[htbp]
	\centerline{\includegraphics[width=1.0\linewidth]{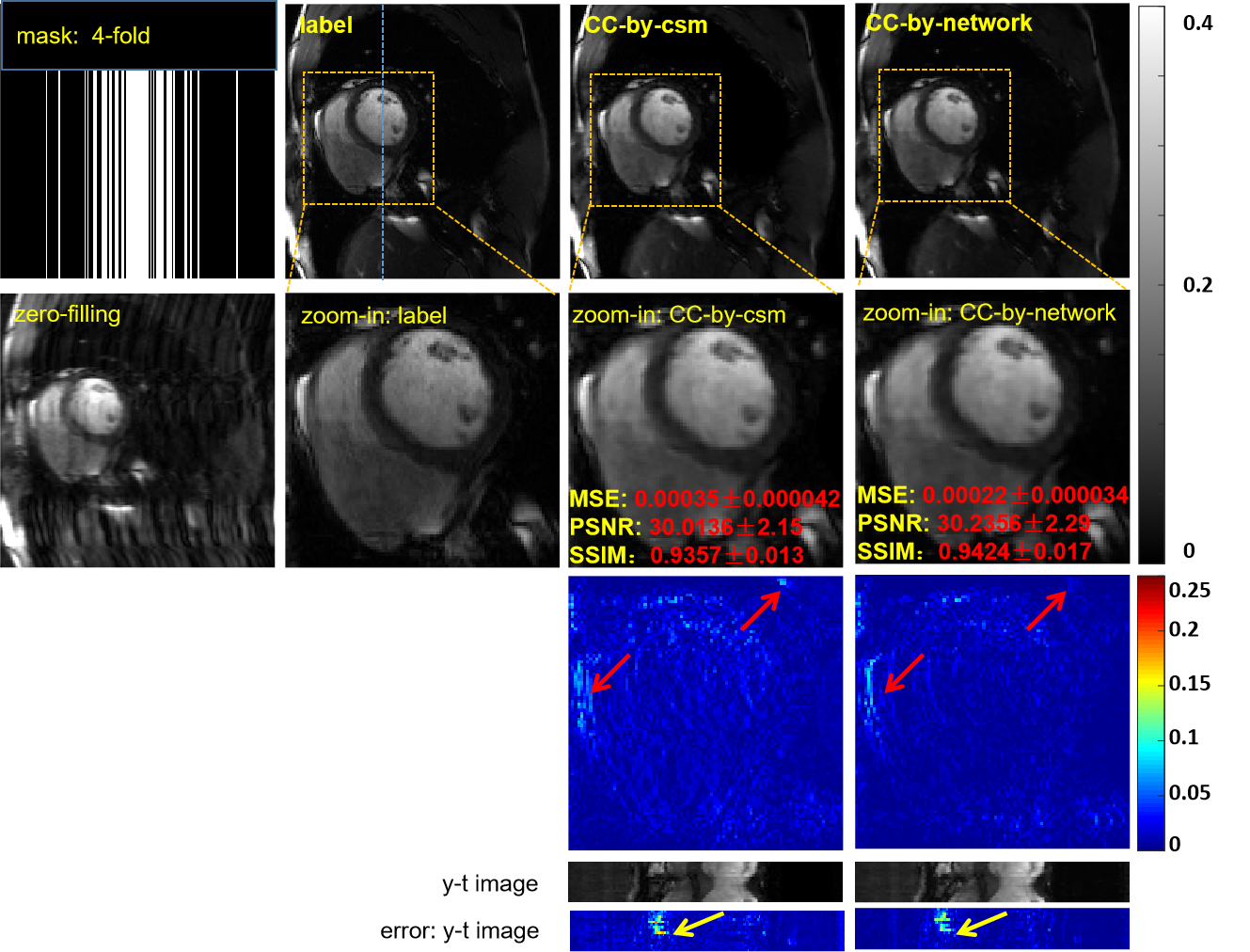}}
	\caption{The reconstruction results of the CC-by-csm and CC-by-network models at 4-fold acceleration. The first row shows, from left to right, the sampling mask, the ground truth, and the reconstruction results of the two models. The second row shows, from left to right, the zero-filling and the enlarged view of their respective heart regions framed by a yellow box. The third row shows the error map (display ranges [0, 0.25]). The y-t image (extraction of the 124th slice along the y and temporal dimensions) and the error of the y-t image are also provided for each signal to show the reconstruction performance in the temporal dimension.\label{fig11}}
\end{figure}

In the proposed method, a coil-combination network was used to explore the coil correlations and combine all the coil images. However, the way to optimally combine coil images via coil sensitivity maps (csm) is well understood. Therefore, the necessity of the coil combination network needs to be discussed. To demonstrate the effectiveness of coil combination by the coil combination network, we built a new model in which its coil combination was completed by csm estimated by ESPIRiT \cite{uecker2014espirit}. The hyperparameters used to estimate csm were selected as follows: calibration region: $24 \times 192$, kernel size: $6 \times 6$, $\sigma^2_{cut-off}= 0.02$, threshold: 0.95. We call this model CC-by-csm and the proposed model CC-by-network. The only difference between the two models is that the coil combination of one model goes through csm and the other through a network. The reconstruction results are shown in Fig.\ref{fig11}. As shown by the error maps, the CC-by-network achieves a better reconstruction performance than does the CC-by-csm model, and its quantitative results are also improved to some extent. Therefore, our coil combination network is helpful in improving the reconstruction results.

\subsection{Training the Model Under Other Sampling Patterns}
\begin{figure*}[htbp]
	\centerline{\includegraphics[width=1.0\linewidth]{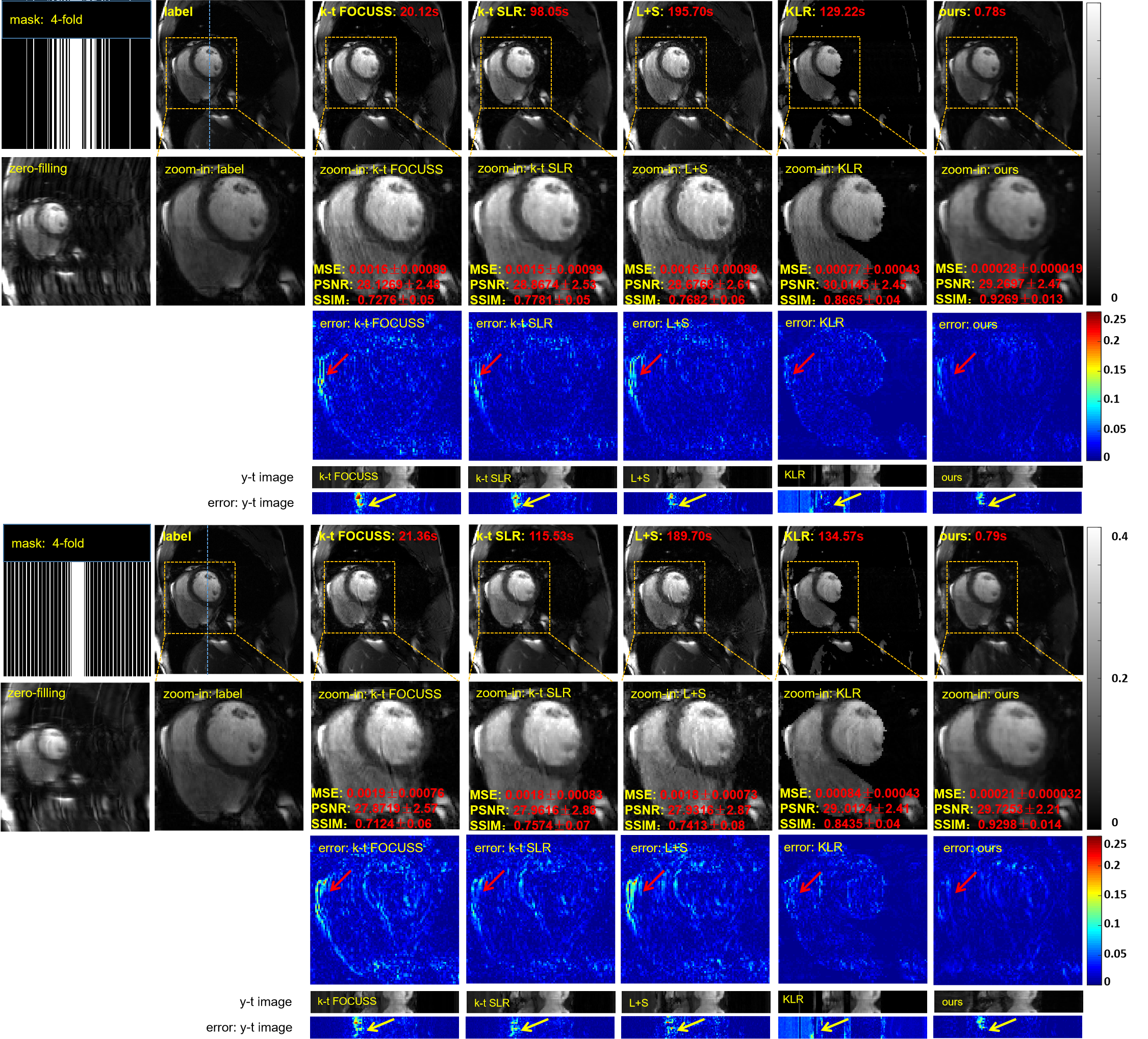}}
	\caption{The reconstruction results of the different methods (k-t FOCUSS, k-t SLR, L+S and the proposed method) at 4-fold acceleration under a 1D random mask and 1D uniform mask. The proposed model is trained under a 1D uniform mask.\label{fig10}}
\end{figure*}

The models in the above sections were all trained with a 1D random sampling pattern. Although our proposed framework is based on a time-interleaved sampling scheme, network training and testing can be conducted with any sampling patterns. The time-interleaved sampling scheme is used only during the data preparation phase. After the fully encoded training data are constructed, retrospective undersampling is no longer dependent on the time-interleaved sampling pattern. Moreover, a model trained on one sampling pattern is well generalized to other sampling patterns. In this section, we trained the model under a 1D uniform undersampling pattern and tested it under 1D random and 1D uniform undersampling patterns. The reconstruction results at 4-fold acceleration are shown in Fig.\ref{fig10}. Our method achieves superior reconstruction results using both undersampling patterns, especially in the heart region, which is marked by the red arrow. 

Although we only show two Cartesian sampling patterns (uniform and random Gaussian) in this section, our method can easily be adopted for other sampling patterns. This finding also reflects one of the advantages of deep learning-based approaches: they have no strict requirements for sampling masks. In another example, Knoll et al \cite{knoll2019assessment} applied a network trained from regular undersampling mask to random undersampling data and achieved good reconstruction performance. Further exploration will be carried out in future studies.

\subsection{The Limitations of the Proposed Work}
Although our method has many advantages (i.e., superior reconstruction results and the shortest reconstruction time) compared with other state-of-the-art methods, the reconstructed images still exhibit a certain degree of oversmoothing. The reason may be our selection of the MSE for the loss function. The MSE loss has a limited ability to perceive image structure information because it indicates only the mean square information between the reconstructed image and the ground truth. DAGAN \cite{yang2017dagan} coupled an adversarial loss with an innovative content loss to reconstruct CS-MRI images, which could preserve perceptual image details. This advantage has motivated us to use more detail-friendly loss functions in future works. 
601 West Main Street Suite 102
Durham, NC 27701 USA

In TGRAPPA \cite{breuer2005dynamic}, more neighboring frames could be averaged to increase the SNR of the fully encoded data. Inspired by this finding, we averaged all the frames to obtain the highest SNR, which yielded some benefits, such as the elimination of temporal redundancies, obtaining better reconstructions and reducing the GPU memory requirement. However, there are some inconveniences. For example, the temporal correlations are underutilized, and many time-dependent network configurations are not available. In the current GPU condition (12 GB memory), averaging all the frames to obtain only one frame is necessary because the GPU resources are insufficient to meet the requirements for exploring the temporal and coil correlations simultaneously. In the future, as the hardware conditions improve, more high-dimensional exploration is expected to further improve dynamic MR image reconstruction. Additionally, the temporal resolution of the built reference data is reduced. However, the temporal resolution of the test data will not changed because no merge operation is required during the test phase.

Breathing patterns vary person to person, which may affect model generalization. However, this problem can be addressed by using a training set that includes richer breathing patterns or using fine-tuning techniques \cite{han2018deep, zhu2018image, Tajbakhsh2016Convolutional}.

\section{CONCLUSION AND OUTLOOK}
In this paper, we propose an unsupervised deep learning method for multi-coil cine MR imaging via time-interleaved sampling. In the framework, fully sampled reference data are no longer required for network training. The temporal redundancies can be effectively utilized via the proposed data preparation process. We also propose a coil-by-coil parallel imaging technology that exhibits many advantages. To the best of our knowledge, this is the first time that a parallel imaging network has been applied to dynamic MR imaging. Although our proposed framework is based on a time-interleaved sampling scheme, the model can be applied to any sampling pattern. The experimental results show that the proposed method is superior to conventional CS-based methods such as k-t FOCUSS, k-t SLR, L+S and KLR in an extremely short amount of time. These findings demonstrate the effectiveness of unsupervised learning and the parallel network in cine MR imaging.

\section*{Acknowledgment}
This research was partly supported by the National Natural Science Foundation of China (61771463, 81830056, U1805261, 81971611, 61871373, 81729003, 81901736); National Key R$\&$D Program of China (2017YFC0108802 and 2017YFC0112903); Natural Science Foundation of Guangdong Province (2018A0303130132); Shenzhen Key Laboratory of Ultrasound Imaging and Therapy (ZDSYS20180206180631473); Shenzhen Peacock Plan Team Program (KQTD20180413181834876); Innovation and Technology Commission of the government of Hong Kong SAR (MRP/001/18X); Strategic Priority Research Program of Chinese Academy of Sciences (XDB25000000).

\ifCLASSOPTIONcaptionsoff
  \newpage
\fi



%

\bibliographystyle{IEEEtran} 
\bibliography{Unsupervised_revision_marked}

\begin{thebibliography}{10}
\providecommand{\url}[1]{#1}
\csname url@samestyle\endcsname
\providecommand{\newblock}{\relax}
\providecommand{\bibinfo}[2]{#2}
\providecommand{\BIBentrySTDinterwordspacing}{\spaceskip=0pt\relax}
\providecommand{\BIBentryALTinterwordstretchfactor}{4}
\providecommand{\BIBentryALTinterwordspacing}{\spaceskip=\fontdimen2\font plus
\BIBentryALTinterwordstretchfactor\fontdimen3\font minus
  \fontdimen4\font\relax}
\providecommand{\BIBforeignlanguage}[2]{{%
\expandafter\ifx\csname l@#1\endcsname\relax
\typeout{** WARNING: IEEEtran.bst: No hyphenation pattern has been}%
\typeout{** loaded for the language `#1'. Using the pattern for}%
\typeout{** the default language instead.}%
\else
\language=\csname l@#1\endcsname
\fi
#2}}
\providecommand{\BIBdecl}{\relax}
\BIBdecl

\bibitem{finn2006cardiac}
J.~P. Finn, K.~Nael, V.~Deshpande, O.~Ratib, and G.~Laub, ``Cardiac {MR}
  imaging: state of the technology,'' \emph{Radiology}, vol. 241, no.~2, pp.
  338--354, 2006.

\bibitem{Liang2007Spatiotemporal}
Z.~P. Liang, ``Spatiotemporal imaging with partially separable functions,'' in
  \emph{Proceedings of 4th IEEE international symposium on biomedical imaging
  (ISBI)}.\hskip 1em plus 0.5em minus 0.4em\relax IEEE, 2007, pp. 988--991.

\bibitem{jerri1977shannon}
A.~J. Jerri, ``The {Shannon} sampling theorem—{Its} various extensions and
  applications: A tutorial review,'' \emph{Proceedings of the IEEE}, vol.~65,
  no.~11, pp. 1565--1596, 1977.

\bibitem{baraniuk2007compressive}
R.~G. Baraniuk, ``Compressive sensing,'' \emph{IEEE signal processing
  magazine}, vol.~24, no.~4, 2007.

\bibitem{lustig2007sparse}
M.~Lustig, D.~Donoho, and J.~M. Pauly, ``Sparse {MRI}: The application of
  compressed sensing for rapid {MR} imaging,'' \emph{Magnetic Resonance in
  Medicine: An Official Journal of the International Society for Magnetic
  Resonance in Medicine}, vol.~58, no.~6, pp. 1182--1195, 2007.

\bibitem{jung2007improved}
H.~Jung, J.~C. Ye, and E.~Y. Kim, ``Improved k--t {BLAST} and k--t {SENSE}
  using {FOCUSS},'' \emph{Physics in Medicine \& Biology}, vol.~52, no.~11, p.
  3201, 2007.

\bibitem{tsao2003k}
J.~Tsao, P.~Boesiger, and K.~P. Pruessmann, ``k-t {BLAST} and k-t {SENSE}:
  dynamic {MRI} with high frame rate exploiting spatiotemporal correlations,''
  \emph{Magnetic Resonance in Medicine: An Official Journal of the
  International Society for Magnetic Resonance in Medicine}, vol.~50, no.~5,
  pp. 1031--1042, 2003.

\bibitem{liang2012k}
D.~Liang, E.~V. DiBella, R.-R. Chen, and L.~Ying, ``k-t {ISD}: dynamic cardiac
  {MR} imaging using compressed sensing with iterative support detection,''
  \emph{Magnetic resonance in medicine}, vol.~68, no.~1, pp. 41--53, 2012.

\bibitem{wang2013compressed}
Y.~Wang and L.~Ying, ``Compressed sensing dynamic cardiac cine {MRI} using
  learned spatiotemporal dictionary,'' \emph{IEEE transactions on Biomedical
  Engineering}, vol.~61, no.~4, pp. 1109--1120, 2013.

\bibitem{otazo2015low}
R.~Otazo, E.~Candes, and D.~K. Sodickson, ``Low-rank plus sparse matrix
  decomposition for accelerated dynamic {MRI} with separation of background and
  dynamic components,'' \emph{Magnetic Resonance in Medicine}, vol.~73, no.~3,
  pp. 1125--1136, 2015.

\bibitem{lingala2011accelerated}
S.~G. Lingala, Y.~Hu, E.~DiBella, and M.~Jacob, ``Accelerated dynamic {MRI}
  exploiting sparsity and low-rank structure: kt {SLR},'' \emph{IEEE
  transactions on medical imaging}, vol.~30, no.~5, pp. 1042--1054, 2011.

\bibitem{Nakarmi2017A}
U.~Nakarmi, Y.~Wang, J.~Lyu, D.~Liang, and L.~Ying, ``A kernel-based low-rank
  ({KLR}) model for low-dimensional manifold recovery in highly accelerated
  dynamic {MRI},'' \emph{IEEE Transactions on Medical Imaging}, vol.~PP,
  no.~99, pp. 1--1, 2017.

\bibitem{ShettyBi2019bi}
G.~N. Shetty, K.~Slavakis, A.~Bose, U.~Nakarmi, G.~Scutari, and L.~Ying,
  ``Bi-linear modeling of data manifolds for dynamic-{MRI} recovery,''
  \emph{IEEE Transactions on Medical Imaging}, 2019.

\bibitem{dong2019deep}
D.~Liang, J.~Cheng, Z.~Ke, and L.~Ying, ``Deep magnetic resonance image
  reconstruction: Inverse problems meet neural networks,'' \emph{IEEE Signal
  Processing Magazine}, 2019.

\bibitem{wang2016accelerating}
S.~Wang, Z.~Su, L.~Ying, X.~Peng, S.~Zhu, F.~Liang, D.~Feng, and D.~Liang,
  ``Accelerating magnetic resonance imaging via deep learning,'' in \emph{2016
  IEEE 13th International Symposium on Biomedical Imaging (ISBI)}.\hskip 1em
  plus 0.5em minus 0.4em\relax IEEE, 2016, pp. 514--517.

\bibitem{kwon2017parallel}
K.~Kwon, D.~Kim, and H.~Park, ``A parallel {MR} imaging method using multilayer
  perceptron,'' \emph{Medical physics}, vol.~44, no.~12, pp. 6209--6224, 2017.

\bibitem{han2018deep}
Y.~Han, J.~Yoo, H.~H. Kim, H.~J. Shin, K.~Sung, and J.~C. Ye, ``Deep learning
  with domain adaptation for accelerated projection-reconstruction {MR},''
  \emph{Magnetic resonance in medicine}, vol.~80, no.~3, pp. 1189--1205, 2018.

\bibitem{zhu2018image}
B.~Zhu, J.~Z. Liu, S.~F. Cauley, B.~R. Rosen, and M.~S. Rosen, ``Image
  reconstruction by domain-transform manifold learning,'' \emph{Nature}, vol.
  555, no. 7697, p. 487, 2018.

\bibitem{eo2018kiki}
T.~Eo, Y.~Jun, T.~Kim, J.~Jang, H.-J. Lee, and D.~Hwang, ``{KIKI}-net:
  cross-domain convolutional neural networks for reconstructing undersampled
  magnetic resonance images,'' \emph{Magnetic resonance in medicine}, vol.~80,
  no.~5, pp. 2188--2201, 2018.

\bibitem{sun2018compressed}
L.~Sun, Z.~Fan, Y.~Huang, X.~Ding, and J.~Paisley, ``Compressed sensing {MRI}
  using a recursive dilated network,'' in \emph{Thirty-Second AAAI Conference
  on Artificial Intelligence}, 2018.

\bibitem{quan2018compressed}
T.~M. Quan, T.~Nguyen-Duc, and W.-K. Jeong, ``Compressed sensing {MRI}
  reconstruction using a generative adversarial network with a cyclic loss,''
  \emph{IEEE transactions on medical imaging}, vol.~37, no.~6, pp. 1488--1497,
  2018.

\bibitem{schlemper2018deep}
J.~Schlemper, J.~Caballero, J.~V. Hajnal, A.~N. Price, and D.~Rueckert, ``A
  deep cascade of convolutional neural networks for dynamic {MR} image
  reconstruction,'' \emph{IEEE Transactions on Medical Imaging}, vol.~37,
  no.~2, pp. 491--503, 2018.

\bibitem{qin2018convolutional}
C.~Qin, J.~V. Hajnal, D.~Rueckert, J.~Schlemper, J.~Caballero, and A.~N. Price,
  ``Convolutional recurrent neural networks for dynamic {MR} image
  reconstruction,'' \emph{IEEE Transactions on Medical Imaging}, vol.~38,
  no.~1, pp. 280--290, 2019.

\bibitem{shan2019dimension}
S.~Wang, Z.~Ke, H.~Cheng, S.~Jia, L.~Ying, H.~Zheng, and D.~Liang,
  ``{DIMENSION}: Dynamic {MR} imaging with both k-space and spatial prior
  knowledge obtained via multi-supervised network training,'' \emph{NMR in
  Biomedicine}, vol. e4131, 2019.

\bibitem{sun2016deep}
J.~Sun, H.~Li, Z.~Xu \emph{et~al.}, ``Deep {ADMM-Net} for compressive sensing
  {MRI},'' in \emph{Advances in neural information processing systems}, 2016,
  pp. 10--18.

\bibitem{hammernik2018learning}
K.~Hammernik, T.~Klatzer, E.~Kobler, M.~P. Recht, D.~K. Sodickson, T.~Pock, and
  F.~Knoll, ``Learning a variational network for reconstruction of accelerated
  {MRI} data,'' \emph{Magnetic resonance in medicine}, vol.~79, no.~6, pp.
  3055--3071, 2018.

\bibitem{cheng2019modellearning}
J.~Cheng, H.~Wang, L.~Ying, and D.~Liang, ``Model learning: Primal dual
  networks for fast {MR} imaging,'' in \emph{International Conference on
  Medical Image Computing and Computer-Assisted Intervention}.\hskip 1em plus
  0.5em minus 0.4em\relax Springer, 2019, pp. 21--29.

\bibitem{aggarwal2018modl}
H.~K. Aggarwal, M.~P. Mani, and M.~Jacob, ``Modl: Model-based deep learning
  architecture for inverse problems,'' \emph{IEEE transactions on medical
  imaging}, vol.~38, no.~2, pp. 394--405, 2018.

\bibitem{cheng2019model}
J.~Cheng, H.~Wang, Y.~Zhu, Q.~Liu, L.~Ying, and D.~Liang, ``Model-based {Deep}
  {MR} {Imaging}: the roadmap of generalizing compressed sensing model using
  deep learning,'' \emph{arXiv preprint arXiv:1906.08143}, 2019.

\bibitem{8962951}
F.~{Knoll}, K.~{Hammernik}, C.~{Zhang}, S.~{Moeller}, T.~{Pock}, D.~K.
  {Sodickson}, and M.~{Akcakaya}, ``Deep-learning methods for parallel magnetic
  resonance imaging reconstruction: A survey of the current approaches, trends,
  and issues,'' \emph{IEEE Signal Processing Magazine}, vol.~37, no.~1, pp.
  128--140, 2020.

\bibitem{knoll2019assessment}
F.~Knoll, K.~Hammernik, E.~Kobler, T.~Pock, M.~P. Recht, and D.~K. Sodickson,
  ``Assessment of the generalization of learned image reconstruction and the
  potential for transfer learning,'' \emph{Magnetic resonance in medicine},
  vol.~81, no.~1, pp. 116--128, 2019.

\bibitem{jun2019parallel}
Y.~Jun, T.~Eo, H.~Shin, T.~Kim, H.-J. Lee, and D.~Hwang, ``Parallel imaging in
  time-of-flight magnetic resonance angiography using deep multistream
  convolutional neural networks,'' \emph{Magnetic resonance in medicine},
  vol.~81, no.~6, pp. 3840--3853, 2019.

\bibitem{akccakaya2019scan}
M.~Ak{\c{c}}akaya, S.~Moeller, S.~Weing{\"a}rtner, and K.~U{\u{g}}urbil,
  ``Scan-specific robust artificial-neural-networks for k-space interpolation
  ({RAKI}) reconstruction: Database-free deep learning for fast imaging,''
  \emph{Magnetic resonance in medicine}, vol.~81, no.~1, pp. 439--453, 2019.

\bibitem{shanshan2019investigation}
S.~Wang, Z.~Ke, H.~Cheng, L.~Ying, X.~liu, H.~Zheng, and D.~Liang,
  ``Investigation of convolutional neural network based deep learning for
  cardiac imaging,'' in \emph{Proceedings of the 26th annual meeting of ISMRM,
  Paris}, no. 2786, 2019.

\bibitem{Fan_2018_ECCV}
Z.~Fan, L.~Sun, X.~Ding, Y.~Huang, C.~Cai, and J.~Paisley, ``A
  segmentation-aware deep fusion network for compressed sensing {MRI},'' in
  \emph{The European Conference on Computer Vision (ECCV)}, September 2018.

\bibitem{el2014content}
K.~El~Asnaoui, B.~Aksasse, and M.~Ouanan, ``Content-based color image retrieval
  based on the 2-d histogram and statistical moments,'' in \emph{2014 Second
  World Conference on Complex Systems (WCCS)}.\hskip 1em plus 0.5em minus
  0.4em\relax IEEE, 2014, pp. 653--656.

\bibitem{lecun2015deep}
Y.~LeCun, Y.~Bengio, and G.~Hinton, ``Deep learning,'' \emph{Nature}, vol. 521,
  no. 7553, pp. 436--444, 2015.

\bibitem{glorot2010understanding}
X.~Glorot and Y.~Bengio, ``Understanding the difficulty of training deep
  feedforward neural networks,'' in \emph{Proceedings of the thirteenth
  international conference on artificial intelligence and statistics}, 2010,
  pp. 249--256.

\bibitem{glorot2011deep}
X.~Glorot, A.~Bordes, and Y.~Bengio, ``Deep sparse rectifier neural networks,''
  in \emph{Proceedings of the fourteenth international conference on artificial
  intelligence and statistics}, 2011, pp. 315--323.

\bibitem{zeiler2012adadelta}
M.~D. Zeiler, ``{ADADELTA}: an adaptive learning rate method,'' \emph{arXiv
  preprint arXiv:1212.5701}, 2012.

\bibitem{kingma2014adam}
D.~P. Kingma and J.~Ba, ``Adam: A method for stochastic optimization,''
  \emph{arXiv preprint arXiv:1412.6980}, 2014.

\bibitem{abadi2016tensorflow}
M.~Abadi, P.~Barham, J.~Chen, Z.~Chen, A.~Davis, J.~Dean, M.~Devin,
  S.~Ghemawat, G.~Irving, M.~Isard \emph{et~al.}, ``Tensorflow: A system for
  large-scale machine learning,'' in \emph{12th Symposium on Operating Systems
  Design and Implementation)}, 2016, pp. 265--283.

\bibitem{walsh2000adaptive}
D.~O. Walsh, A.~F. Gmitro, and M.~W. Marcellin, ``Adaptive reconstruction of
  phased array {MR} imagery,'' \emph{Magnetic Resonance in Medicine: An
  Official Journal of the International Society for Magnetic Resonance in
  Medicine}, vol.~43, no.~5, pp. 682--690, 2000.

\bibitem{uecker2014espirit}
M.~Uecker, P.~Lai, M.~J. Murphy, P.~Virtue, M.~Elad, J.~M. Pauly, S.~S.
  Vasanawala, and M.~Lustig, ``{ESPIRiT}—an eigenvalue approach to
  autocalibrating parallel {MRI}: where {SENSE} meets {GRAPPA},''
  \emph{Magnetic resonance in medicine}, vol.~71, no.~3, pp. 990--1001, 2014.

\bibitem{yang2017dagan}
G.~Yang, S.~Yu, H.~Dong, G.~Slabaugh, P.~L. Dragotti, X.~Ye, F.~Liu,
  S.~Arridge, J.~Keegan, Y.~Guo \emph{et~al.}, ``{DAGAN}: deep de-aliasing
  generative adversarial networks for fast compressed sensing {MRI}
  reconstruction,'' \emph{IEEE transactions on medical imaging}, vol.~37,
  no.~6, pp. 1310--1321, 2017.

\bibitem{breuer2005dynamic}
F.~A. Breuer, P.~Kellman, M.~A. Griswold, and P.~M. Jakob, ``Dynamic
  autocalibrated parallel imaging using temporal {GRAPPA (TGRAPPA)},''
  \emph{Magnetic Resonance in Medicine: An Official Journal of the
  International Society for Magnetic Resonance in Medicine}, vol.~53, no.~4,
  pp. 981--985, 2005.

\bibitem{Tajbakhsh2016Convolutional}
N.~Tajbakhsh, J.~Y. Shin, S.~R. Gurudu, R.~T. Hurst, C.~B. Kendall, M.~B.
  Gotway, and J.~Liang, ``Convolutional neural networks for medical image
  analysis: Full training or fine tuning?'' \emph{IEEE transactions on medical
  imaging}, vol.~35, no.~5, pp. 1299--1312, 2016.

\end{thebibliography}

%






\end{document}